\documentclass[hyper]{JHEP} 

\usepackage{epsfig}

\newcommand\fverb{\setbox\pippobox=\hbox\bgroup\verb}

\newcommand\fverbdo{\egroup\medskip\noindent%

            \fbox{\unhbox\pippobox}\ }

\newcommand\fverbit{\egroup\item[\fbox{\unhbox\pippobox}]}

\newbox\pippobox

\title{Hamiltonian Analysis of
Non-Relativistic Covariant RFDiff
Ho\v{r}ava-Lifshitz Gravity}
\author{J. Kluso\v{n}\\
Department of
Theoretical Physics and Astrophysics\\
Faculty of Science, Masaryk University\\
Kotl\'{a}\v{r}sk\'{a} 2, 611 37, Brno\\
Czech Republic\\
E-mail: \email{klu@physics.muni.cz}}
\preprint{}

 \abstract{We perform the Hamiltonian
 analysis
 of  non-relativistic covariant Ho\v{r}ava-Lifshitz
 gravity in the formulation presented recently in
 arXiv:1009.4885. We argue that the resulting Hamiltonian
 structure is in agreement with the original
 construction of non-relativistic covariant Ho\v{r}ava-Lifshitz
 gravity presented  in arXiv:1007.2410.
 Then we extend this construction  to the case
   of  RFDiff invariant Ho\v{r}ava-Lifshitz theory. We
 find well behaved Hamiltonian system
 with the number of the first and the second class constraints
 that ensure the correct  number of physical degrees of freedom
 of gravity.} 

\def\be{\begin{equation}}

\def\ee{\end{equation}}

\def\bea{\begin{eqnarray}}

\def\eea{\end{eqnarray}}

\def\tK{\tilde{K}}
\def\hN{\hat{N}}
\def\hK{\hat{K}}

\def\mH{\mathcal{H}}

\def\bz{\mathbf{z}}

\def\bx{\mathbf{x}}
\def\by{\mathbf{y}}

\newcommand{\mA}{\mathcal{A}}

\newcommand{\mG}{\mathcal{G}}

\def\mV{\mathcal{V}}

\newcommand{\bT}{\mathbf{T}}

\def\pb #1{\left\{#1\right\}}

\begin{document}
\section{Introduction and Summary}\label{first}
In 2009 Petr Ho\v{r}ava
formulated new proposal of
quantum theory of gravity that
is power counting renormalizable
\cite{Horava:2009uw,Horava:2008ih,Horava:2008jf}.
This theory is now known as Ho\v{r}ava-Lifshitz
gravity (HL gravity).
It was also expected that
this theory reduces do General
Relativity in the infrared (IR)
limit. HL theory was studied
from different point of view
due to the fact that this is
 a new and intriguing formulation
of gravity as a theory with reduced
amount of symmetries that leads
to remarkable new phenomena
\footnote{For review and extensive
list of references, see
\cite{Padilla:2010ge,Mukohyama:2010xz,Weinfurtner:2010hz}.}.

The HL gravity is based on an idea that
the Lorentz symmetry is restored in IR
limit of given theory and can be absent
at high energy regime of given theory.
Explicitly,  Ho\v{r}ava considered
systems whose scaling at short
distances exhibits a strong anisotropy
between space and time,
\begin{equation}
\bx' =l \bx \ , \quad t' =l^{z} t \ .
\end{equation}
In $(D+1)$ dimensional space-time in
order to have power counting
renormalizable theory  requires that
$z\geq D$. It turns out however that
the symmetry group of given theory is
reduced from the full diffeomorphism
invariance of General Relativity  to
the foliation preserving diffeomorphism
\begin{equation}\label{fpdi}
x'^i=x^i+\zeta^ i(t,\bx) \ , \quad
t'=t+f(t) \ .
\end{equation}
Due to the fact that the diffeomorphism
is restricted (\ref{fpdi})  one more
degree of freedom appears that is a
spin$-0$ graviton. It turns out that
the existence of this mode could be
dangerous since it has to decouple
 in the IR regime, in order to
be consistent with observations.
Unfortunately, it seems  that this
might not be the case. It was shown
that the spin-0 mode is not stable in
the original version of the HL theory
[1] as well as in the Sotiriou, Visser
and Weinfurtner (SVW) generalization
\cite{Sotiriou:2009bx}. Note that in
both of these two versions, it was all
assumed the projectability condition
that means that the lapse function $N$
depends on $t$ only. This presumption
has a fundamental consequence for the
formulation of the theory since there
is no local form of the Hamiltonian
constraint but the only global one.
However we would like to stress that
these instabilities are all found in
the Minkowski background. Recently, it
was found that the de Sitter spacetime
is stable in the SVW setup
\cite{Huang:2010rq,Wang:2010uga}. Then
we can presume that this background is
legitimate background.

On the other hand there is the second
version of HL gravity where the
projectability condition is not imposed
so that $N=N(\bx,t)$. Properties of
given theory were extensively studied
in
\cite{Blas:2009yd,Blas:2009qj,Blas:2009ck,Blas:2010hb,Li:2009bg,Henneaux:2009zb,
Kluson:2010xx,Kluson:2010nf,Bellorin:2010te,Bellorin:2010je,Kobakhidze:2009zr,Pons:2010ke}.
It was shown recently in
\cite{Blas:2010hb} that  so called
healthy extended version of given
theory could really be an interesting
candidate for the quantum theory of
reality without ghosts and without
strong coupling problem despite its
unusual Hamiltonian structure
\cite{Kluson:2010xx,Kluson:2010nf}.

Recently Ho\v{r}ava and Malby-Thompson
in \cite{Horava:2010zj}
 proposed very interesting way how
to eliminate the spin-0 graviton. They
considered the projectable version of
HL gravity together with extension of
the foliation preserving diffeomorphism
to include a local $U(1)$ symmetry. The
resulting theory is then called as
non-relativistic  covariant theory of
gravity \footnote{This theory was also
studied in
\cite{Greenwald:2010fp,Alexandre:2010cb,Wang:2010wi,Huang:2010ay}.}.
It was argued there
\cite{Horava:2010zj} that the presence
of this new symmetry forces the
coupling constant $\lambda$ to be equal
to one, however this result was
questioned in \cite{daSilva:2010bm}
(see also \cite{Huang:2010ay}) where an
alternative formulation of
non-relativistic general covariant
theory of gravity was presented.
Further, it was shown in
\cite{Horava:2010zj,daSilva:2010bm}
that the presence of this new symmetry
implies that the spin-0 graviton
becomes non-propagating and the
spectrum of the linear fluctuations
around the background solution
coincides with the fluctuation spectrum
of General Relativity.

This new proposal of non-relativistic
general covariant HL gravity is very
interesting and it certainly deserves
further study. In this paper we present
the Hamiltonian analysis of the
formulation of non-relativistic
covariant HL gravity  given in
\cite{daSilva:2010bm}. We argue that
resulting Hamiltonian and constraint
structure has the same form as in
\cite{Horava:2010zj} even if they
differ in explicit form since they are
derived from different Lagrangians.
This fact shows that these two
Lagrangian formulations of
non-relativistic covariant HL gravities
are equivalent on the level of the
Hamiltonian formalism as well.

Despite the fact that non-relativistic
covariant HL gravity seems to solve the
problem of the  scalar graviton and the
content of the physical degrees of
freedom is the same as in General
Relativity there is still one
additional first class constraint which
is the global Hamiltonian constraint.
The  meaning of this constraint should
be investigated further as was nicely
discussed on page 30 in
\cite{Horava:2010zj}. In order to
 find   version of non-relativistic
covariant HL gravity without global
Hamiltonian constraint   we recall that
there exists formulation of the HL
gravity with reduced symmetry group
known as
\emph{restricted-foliation-preserving
Diff} (RFDiff)
 HL gravity
\cite{Blas:2010hb,Kluson:2010na}. This
is the theory that is invariant under
following symmetries
\begin{equation}
t'=t+\delta t \ , \delta
t=\mathrm{const} \ , \quad x'^i=
x^i+\zeta^i(\bx,t) \ .
\end{equation}
The characteristic property of given
theory is that in its simplest version
\cite{Kluson:2010na} based on the
detailed balance construction
\cite{Horava:2009uw,Horava:2008ih,Horava:2008jf}
there is no reason to introduce the
lapse function $N$ \footnote{More
general form of RFDiff HL gravity was
considered in \cite{Blas:2010hb} where
the action contains time and space
derivatives of the lapse function $N$
according to general principles of
effective field theory construction.
 However the
presence of such terms has no impact on
the Hamiltonian structure of given
theory simply from the fact that the
momentum conjugate to $N$ is not
primary constraint of the theory and
hence the Hamiltonian constraint is
absent. In order to make our analysis
transparent   we consider the simplest
version of RFDiff HL gravity keeping in
mind that it can be easily extended to
its more general versions.}. Then we
introduce $U(1)$ symmetry as in
\cite{Horava:2010zj} or its alternative
version given in \cite{daSilva:2010bm}.
Finally we proceed to the Hamiltonian
formulation of given theory and we find
there is no global Hamiltonian
constraint due to the absence of the
lapse function $N$ in the action. We
further determine all constraints in
given theory and we show that the
number of the  first class and the
second class constraints implies that
the physical phase space has the same
dimensions as in case of General
Relativity. On the other hand we show
that the presence of the second class
constraints implies that the symplectic
structure of given theory that is
determined by  corresponding Dirac
brackets between physical degrees of
freedom  is rather complicated due to
the fact that  Dirac brackets generally
depend on phase space variables.

Let us outline our results and suggest
possible extension of this work. We
perform  the Hamiltonian analysis  of
the theory suggested in
\cite{daSilva:2010bm} and we show that
its Hamiltonian structure is equivalent
to the Hamiltonian structure found in
paper \cite{Horava:2010zj}. We also
suggest an alternative formulation of
non-relativistic covariant HL gravity
that is based RFDiff HL gravity. We
show that the resulting theory has
consistent Hamiltonian formulation with
the same content of the local
constraints as in case of
non-relativistic covariant HL gravity
but without global Hamiltonian
constraint.
 On the other hand we
should stress that we are not able to
solve explicitly the second class
constraints with respect to physical
degrees of freedom in the full
generality. We are also not able to
determine corresponding Dirac brackets.
Then it would be clearly desirable to
find exact  results at least for some
special situations. It would be also
interesting to find exact solutions of
the equations of motion of
non-relativistic general covariant
RFDiff HL gravity. We hope to return to
these problems in near future.

The organization of this paper is as follows.
In the next section (\ref{second}) we introduce
the non-relativistic general covariant HL
gravity in the formulation firstly presented in
\cite{daSilva:2010bm}. Then in section (\ref{third})
we perform its Hamiltonian analysis. In section
(\ref{fourth}) we introduce
the non-relativistic general covariant RFDiff-invariant
HL gravity. Then we perform its Hamiltonian
analysis and shows that the resulting theory correctly
describes physical degrees of freedom of $D+1$ dimensional
gravity.

\section{Non-Relativistic
  Covariant  HL Gravity}\label{second}
We begin  this section with the
introduction of basic notation, for
detailed treatment of $D+1$ formalism, see
\cite{Gourgoulhon:2007ue}.

Let us consider $D+1$ dimensional
manifold $\mathcal{M}$ with the
coordinates $x^\mu \ , \mu=0,\dots,D$
and where $x^\mu=(t,\bx) \ ,
\bx=(x^1,\dots,x^D)$. We presume that
this space-time is endowed with the
metric $\hat{g}_{\mu\nu}(x^\rho)$ with
signature $(-,+,\dots,+)$. Suppose that
$ \mathcal{M}$ can be foliated by a
family of space-like surfaces
$\Sigma_t$ defined by $t=x^0$. Let
$g_{ij}, i,j=1,\dots,D$ denotes the
metric on $\Sigma_t$ with inverse
$g^{ij}$ so that $g_{ij}g^{jk}=
\delta_i^k$. We further introduce the operator
$\nabla_i$ that is covariant derivative
defined with the metric $g_{ij}$.
 We  introduce  the
future-pointing unit normal vector
$n^\mu$ to the surface $\Sigma_t$. In
ADM variables we have
$n^0=\sqrt{-\hat{g}^{00}},
n^i=-\hat{g}^{0i}/\sqrt{-\hat{g}^{
00}}$. We also define  the lapse
function $N=1/\sqrt{-\hat{g}^{00}}$ and
the shift function
$N^i=-\hat{g}^{0i}/\hat{g}^{00}$. In
terms of these variables we write the
components of the metric
$\hat{g}_{\mu\nu}$ as
\begin{eqnarray}
\hat{g}_{00}=-N^2+N_i g^{ij}N_j \ ,
\quad \hat{g}_{0i}=N_i \ , \quad
\hat{g}_{ij}=g_{ij} \ ,
\nonumber \\
\hat{g}^{00}=-\frac{1}{N^2} \ , \quad
\hat{g}^{0i}=\frac{N^i}{N^2} \ , \quad
\hat{g}^{ij}=g^{ij}-\frac{N^i N^j}{N^2}
\ .
\nonumber \\
\end{eqnarray}
Let us now consider the
general form of
Ho\v{r}ava-Lifshitz action
\begin{equation}\label{actionNOJI}
S= \frac{1}{\kappa^2} \int dt d^D\bx
\sqrt{g}N \left[ K_{ij}\mG^{ijkl}K_{kl}
-\mathcal{V}(g)\right]
 \ ,
\end{equation}
where $K_{ij}$ denotes
  the extrinsic
derivative
\begin{equation}
K_{ij}=\frac{1}{2N}
(\partial_t g_{ij}-\nabla_i N_j-
\nabla_j N_i) \ .
\end{equation}
Further the generalized De Witt metric
$\mG^{ijkl}$  is defined as
\begin{equation}
\mG^{ijkl}=\frac{1}{2}(g^{ik}g^{jl}+
g^{il}g^{jk})-\lambda g^{ij}g^{kl} \ ,
\end{equation}
where $\lambda$ is a real constant that in
case of General Relativity is equal to one.
Finally $\mV(g)$ is general function
of $g_{ij}$ and its covariant derivative.
Note also that we consider \emph{projectable
} version of HL gravity where $N=N(t)$.

The action
(\ref{actionNOJI})
 is invariant under foliation
preserving diffeomorphism
\begin{equation}\label{fpd}
t'-t=f(t) \ , \quad
x'^i-x^i=\xi^i(t,\bx) \ .
\end{equation}
Following \cite{Horava:2010zj}
we introduce $U(1)$ transformation
with parameter $\alpha(\bx,t)$ under
which $g_{ij},N_i$ and $N$ transform
as
\begin{equation}\label{deltaNi}
\delta_\alpha N=0 \ , \quad, \delta_\alpha g_{ij}(\bx,t)=0 \ ,
\quad \delta_\alpha N_i(\bx,t)=
N(t)\nabla_i\alpha(\bx,t) \ .
\end{equation}
As was shown in \cite{Horava:2010zj}
the action (\ref{actionNOJI}) is not
invariant under the transformation
(\ref{deltaNi}) at least for $D\neq 2$.
Then the general procedure how to find
an invariant action was formulated in
\cite{Horava:2010zj}. It is based on an
introducing of the scalar field $\nu$
that transforms under (\ref{deltaNi})
as
\begin{equation}
\delta_\alpha \nu(t,\bx)=
\alpha(t,\bx) \ .
\end{equation}
Then it turns out that the action
invariant under (\ref{deltaNi}) can be
written in the form
\begin{eqnarray}\label{SFtR}
S
=\frac{1}{\kappa^2}\int dt d^D\bx
\sqrt{g}N (
(K_{ij}+\nabla_i\nabla_j\nu)\mG^{ijkl}(K_{kl}+\nabla_k\nabla_l
\nu) -\mV(g))
\end{eqnarray}
or in even more suggestive form by introducing
\begin{equation}
\hN_i=N_i-N\nabla_i \nu  \ ,
\quad
\hK_{ij}=\frac{1}{2N}(\partial_t g_{ij}
-\nabla_i \hN_j-\nabla_j \hN_i)
\end{equation}
so that
\begin{eqnarray}\label{SFtRfinal}
S =\frac{1}{\kappa^2}\int dt d^D\bx
 \sqrt{g}N \left[
\hK_{ij}\mG^{ijkl}\hK_{kl} -\mV(g)\right] \ .
\end{eqnarray}
 However from this
analysis it is  clear that $\nu$ has a
character of the St\"{u}ckelberg field
and hence the symmetry (\ref{deltaNi})
is trivial. The novelty of the analysis
\cite{Horava:2010zj} in the formulation
\cite{daSilva:2010bm}
 is in the
introduction of the additional term
into action
\begin{equation}\label{Snuk}
S_{\nu,k}= \frac{1}{\kappa^2}\int dt
d^D\bx  \sqrt{g} \mG(g_{ij})(\mA-a) \ ,
\end{equation}
where
\begin{equation}
a=\dot{\nu}-N^i\nabla_i\nu+
\frac{N}{2}
\nabla^i \nabla_i \nu \ .
\end{equation}
In the original work
\cite{Horava:2010zj} the function
$\mG(g)$ was equal to $R-\Omega$ where
$R$ is $D-$dimensional curvature and
$\Omega$ is constant. Note that in
principle it is possible  to consider
more general form of $\mG$ as was
suggested in \cite{daSilva:2010bm}.
Further,  $a$ transforms under $\alpha$
variation as
\begin{eqnarray}
a'(t,\bx)
=a(t,\bx)+\dot{\alpha}(t,\bx)-N^i(t,\bx)
\nabla_i\alpha(t,\bx) \ .  \nonumber \\
\end{eqnarray}
Now when we  presume that $\mA$
transforms under $\alpha$ variation as
\begin{equation}\label{mAtr}
\mA'(t,\bx)=\mA(t,\bx)+\dot{\alpha}(t,\bx)-
N^i(t,\bx)\nabla_i \alpha(t,\bx)
\end{equation}
we immediately find  that (\ref{Snuk})
is invariant under $\alpha-$variation.
Say differently, $\mA$ can be
interpreted as the gauge field that has
to be introduced when we gauge the
$\alpha$ transformation
\cite{Horava:2010zj}. More precisely,
it is clear that
 the action (\ref{SFtRfinal})
is invariant under general
$\alpha(t,\bx)$ however as  we argued
this is  trivial St\"{u}ckelberg
extension with no impact on physical
content of given theory. On the other
hand let us presume that we want to
construct more interesting modification
of given theory when we add
  (\ref{Snuk}) without $\mA$ to the original
  HL action. Now this term is invariant
  under
 $\alpha-$variation on condition that $\alpha$
obeys the equation
\begin{equation}
\dot{\alpha}(t,\bx)-N^i(t,\bx)
\nabla_i\alpha(t,\bx)=0 \ .
\end{equation}
that means that $\alpha$ is covariantly
constant \cite{Horava:2010zj} and hence
should be interpreted as a parameter of
a global symmetry. Gauging this symmetry means
that we relax this condition and also
introduce the gauge field $\mA$ that
transforms as (\ref{mAtr}).

It is clear from the analysis given
above that the non-relativistic
covariant HL gravity is invariant under
(\ref{deltaNi}) for arbitrary $\lambda$
as was firstly stressed in
\cite{daSilva:2010bm}. Then it was
argued that there is no scalar graviton
in the perturbative spectrum about the
flat background that makes this action
very attractive since it solves the
main issue of HL gravity.
\section{Hamiltonian Formalism For Non-relativistic
Covariant HL Gravity}\label{third} For
reader's convenience we  again write
non-relativistic  covariant  HL action
\begin{eqnarray}\label{SFtRFinal2}
S &=&\frac{1}{\kappa^2}\int dt d^D\bx
 \sqrt{g}N (
\hK_{ij}\mG^{ijkl}\hK_{kl}
-\mV(g))+\nonumber
\\
&+&\frac{1}{\kappa^2}\int dt d^D\bx
\sqrt{g} \mG(R)(\mA-a) \ ,
\end{eqnarray}
where we now restrict to the case when
$\mG$ depends $g_{ij}$ through the $D-$
dimensional curvature $R(g_{ij})$
\footnote{We should stress one
important issue that is related to the
form of the action (\ref{SFtRFinal2}).
At present it is not completely clear
how the matter fields should be
included in the action
(\ref{SFtRFinal2}). Clearly we can
trivially couple any matter field with
$\nu$ when we replace $N_i\rightarrow
N_i-N\nabla_i \nu$ in all expressions
containing $N_i$ and given matter
field. On the other hand it is an open
problem how the presence of the matter
field is related to the scalar
curvature. In fact, (\ref{SFtRFinal2})
implies that the scalar curvature is
determined solely by the function
$\mG(R)=0$ that does not depend on the
matter fields. One possibility is to
consider the case when $\mG$ depends on
$(R-g^{ij}\partial_i\phi\partial_j\phi)$
instead of $R$ when the matter field is
represented by the scalar field $\phi$.
The detailed analysis of the action
with given modification could be very
interesting.}.
 From (\ref{SFtRFinal2})
we find the conjugate momenta
\begin{eqnarray}
\pi^{ij}&=& \frac{1}{\kappa^2}\sqrt{g}\mG^{ijkl}\hK_{kl}
\ , \quad p_N\approx 0 \ , \quad p^i
\approx 0 \ ,
\nonumber \\
p_\mA &\approx& 0 \ , \quad
p_\nu=-\frac{1}{\kappa^2}\sqrt{g}\mG \
\nonumber \\
\end{eqnarray}
that imply the $3+D$
primary constraints
\begin{equation}
p_N\approx 0 \ , \quad p^i(\bx)\approx
0 \ , \quad  \Phi_1(\bx):
p_\mA(\bx)\approx 0 \ , \quad
\Phi_2(\bx):
p_\nu(\bx)+\frac{1}{\kappa^2}\sqrt{g}
\mG(\bx)\approx 0  \ .
\end{equation}
Then following standard procedure we
determine the  Hamiltonian in the form
\begin{eqnarray}
H&=&\int d^D\bx (N\mH_T+N^i\mH_i+
v^A\Phi_A+v_Np_N+v_ip^i)-\nonumber \\
&-&\frac{1}{\kappa^2} \int d^D\bx
\sqrt{g}\mG(R) (\mA-N^i \nabla_i
\nu+\frac{N}{2} \nabla^i \nabla_i\nu) \
,
\nonumber \\
\end{eqnarray}
where $v_N,v_i,v^A,A=1,2$ are
Lagrange multipliers related to
corresponding primary constraints
and where
\begin{eqnarray}\label{mHTA}
\mH_T&=&\frac{\kappa^2}{\sqrt{g}}\pi^{ij}\mG_{ijkl}\pi^{kl}
-\frac{1}{\kappa^2}\sqrt{g}\mV(g) -\frac{2}{\kappa^2} \nu \nabla_i\nabla_j \pi^{ij}\ ,  \nonumber \\
\mH_i&=&-2g_{il} \nabla_k \pi^{kl}
 \ .  \nonumber \\
\end{eqnarray}
 Note that
$N$ and $p_N$ do not depend on $\bx$.
 Now
the requirement of the preservation of
the primary constraints $p_N \approx 0,
p_i(\bx)\approx 0 ,
\Phi_{1}(\bx)\approx 0$ implies following
secondary ones
\begin{eqnarray}
\partial_t \Phi_1&=&\pb{\Phi_1,H}=
-\frac{1}{\kappa^2}
\sqrt{g}\
\mG\equiv -
\Phi_1^{II}
\approx 0 \ ,
\nonumber \\
\partial_t p_N&=&\pb{p_N,H}=
 -\int d^D\bx \mH_T
+\frac{1}{2}
\int d^D\bx \Phi_1^{II}
\nabla^i \nabla_i\nu
\approx \nonumber \\
&\approx &
 -\int d^D\bx \mH_T\approx 0
\nonumber \\
\partial_t  p_i&=&\pb{p_i,H}=
-\mH_i- \Phi_1^{II} \approx
-\mH_i\approx 0 \ .
 \  \nonumber \\
 \end{eqnarray}
Now using following formulas
\begin{eqnarray}
\pb{R(\bx),\pi^{ij}(\by)}&=&
-R^{ij}(\bx)\delta(\bx-\by)+ \nabla^i
\nabla^j \delta(\bx-\by)-g^{ij}
\nabla_k \nabla^k\delta(\bx-\by) \ ,
\nonumber \\
\nabla^i \nabla^j \mG_{ijkl}
\pi^{kl}&-& g^{ij}\nabla_m\nabla^m
\mG_{ijkl}\pi^{kl} =\nabla_k (\nabla_l
\pi^{kl}) +\frac{1-\lambda}{\lambda
D-1}\nabla_i
\nabla^i \pi \  \nonumber \\
\end{eqnarray}
we find that the time derivative of
$\Phi_2$ is equal to
 \begin{eqnarray}\label{tPhi2}
\partial_t \Phi_2&=&
\pb{\Phi_2,H}\approx -2N
\frac{d\mG}{dR} \left(
 R^{ij}
\mG_{ijkl}\pi^{kl} -
\frac{1-\lambda}{(\lambda D-1)}\nabla_k
\nabla^k \pi\right)=2N\frac{d\mG}{dR}
\Phi_2^{II} \ ,
\nonumber \\
\end{eqnarray}
where
\begin{eqnarray}
\Phi_2^{II}
=- R_{ij}\pi^{ji}
+\frac{\lambda}{D\lambda-1}R\pi
+ \frac{1-\lambda}{(\lambda
D-1)}\nabla_k \nabla^k \pi
\equiv
M_{ij}(g(\bx))\pi^{ji}(\bx)
\ ,  \nonumber \\
\end{eqnarray}
 where generally $M_{ij}(g(\bx))$ is   a
differential operator acting on $\pi^{ij}$
 that it reduces to ordinary multiplicative
operator in case $\lambda=1$. Note that
in the calculation of (\ref{tPhi2}) we
used following result
\begin{equation}\label{pnuH}
\pb{p_\nu,H}=
-N\nabla^i\mH_i+
\frac{1}{\kappa^2}\nabla_i (\sqrt{g} N^i
\mG)+\frac{N}{2\kappa^2}\nabla^i
\nabla_i (\sqrt{g}\mG)
\approx 0 \ , \nonumber \\
\end{equation}
where in the final step we used the
fact that the result is proportional to
the constraints $\mH_i$ and
$\Phi_1^{II}\approx 0$. In the same way
we find that
\begin{eqnarray}
&-&2\pb{\sqrt{g}\mG,
\int d^D\bx N\nu \nabla_i\nabla_j\pi^{ij}(\bx))}\approx
\nonumber \\
&\approx  &\pb{\int d^D\bx N
\nabla_i\nu \mH^i ,\sqrt{g}\mG}\approx
\sqrt{g}\partial_i \mG \nabla^i\nu
\approx 0 \ . \nonumber \\
\end{eqnarray}

Let us review  constraints that
we derived at this stage. We have following set
of secondary constraints
$\Phi_1^{II}\approx 0 \ , \Phi_2^{II}\approx 0 \ ,  \mH_i\approx
0$ and one global $\bT=\int d^D\bx
\mH_T\approx 0$. Note also that $p_\nu=\Phi_1-\Phi_1^{II}\approx 0$
that according to (\ref{pnuH}) is the first class
constraint.  Then the total Hamiltonian takes the form
\begin{eqnarray}
H_T&=&\int d^D\bx (N\mH_T+N^i\mH_i+
v^\mA p_\mA+v_Np_N+v_ip^i+v^\nu p_\nu+
v^1_{II}\Phi^{II}_1+v^2_{II}
\Phi^{II}_2) \ , \nonumber \\
\end{eqnarray}
where
$v_N,v_i,v^{\mA},v^1_{II},v^2_{II}$ are
corresponding Lagrange multipliers.
Note that we included the expression
$(\mA-N^i\nabla_i\nu+\frac{N}{2}
\nabla^i\nabla_i\nu)$ into definition
of the Lagrange multiplier $v_{II}^1$.

As the final step we analyze the stability of the
secondary constraints.
%
Let us begin with the constraint $\mH_i$.
It is convenient to
extend these constraints by appropriate
combinations of  additional constraints
$p_\nu\approx 0,p_\mA\approx 0$ so that
\begin{equation}
\mH_i=-2g_{ik}\nabla_l\pi^{kl}
+\partial_i \mA p_\mA+
\partial_i \nu p_\nu \ .
\end{equation}
Then $\bT_S(N^i)= \int d^D\bx N^i\mH_i$
is generator of the spatial
diffeomorphism  that is clearly
preserved during the time evolution of
the system since the Hamiltonian is
invariant under spatial diffeomorphism.
Further, $p_\nu\approx 0$ is preserved
during the time evolution of the system
according to (\ref{pnuH}). On the other
hand the time evolution of the
constraint $\Phi_1^{II}\approx 0$ is
equal
to
\begin{eqnarray}\label{parttPhi2}
\partial_t \Phi_1^{II}&=&\pb{\Phi_1^{II},H_T}
\approx  \int d^D\bx \left(N
\sqrt{g}\frac{d\mG}{dR}
\Phi_2^{II}(\bx) +v_{II}^2(\bx)
\pb{\Phi_1^{II},\Phi_2^{II}(\bx)}\right)
\approx \nonumber \\
&\approx& \int d^D\bx
v_{II}^2(\bx)\pb{\Phi_1^{II},\Phi_2^{II}(\bx)}=0
\ .
\nonumber \\
\end{eqnarray}
Since
\begin{eqnarray}\label{DBphi12}
&
&\pb{\Phi_1^{II}(\bx),\Phi_2^{II}(\by)}=
M_{ij}(\by)\left(
\pb{\Phi^{II}_1(\bx),\pi^{ji}(\by)}\right)\approx
\nonumber \\
&\approx&
M_{ij}(\by)\left(\sqrt{g}\frac{\delta
\mG}{\delta R} \frac{\delta
R(\bx)}{\delta
g_{ij}(\by)}\right)
\neq 0 \ . \nonumber \\
\end{eqnarray}
we find that the equation
(\ref{parttPhi2}) implies that
$v_{II}^2=0$. In the same way the
requirement of the preservation of the
constraint $\Phi_2^{II}$ implies
\begin{eqnarray}
\partial_t\Phi_2=
N\pb{\Phi_2,\bT}+\int d^D\bx
v_{II}^1(\bx)\pb{\Phi_2,\Phi_1(\bx)}=0
\nonumber \\
\end{eqnarray}
that due to the fact that
$\pb{\Phi_2,\bT}\neq 0$ and
(\ref{DBphi12}) allows to determine
$v_{II}^1$ as a function of the
canonical variables. In other words,
$\Phi_1^{II}$ and $\Phi_2^{II}$ are the
second class constraints \footnote{The
nontrivial property of given theory is
that the Poisson bracket between the
second class constraints depend on the
phase space variables so that it is
possible that it vanishes on some
subspace of phase space. In order to
analyze this issue we should explicitly
determine the form of this Poisson
bracket and after some algebra we find
$\pb{\Phi_1^{II}(\bx),\Phi_2^{II}(\by)}=
\mathbf{\triangle}(R,R_{ij},\bx,\by)+
\frac{(1-\lambda)(1-D)}{D\lambda-1}
\nabla_i\nabla^i\nabla_j\nabla^j
\delta(\bx-\by)$. We see that due to
the second therm this Poisson bracket
is non-zero on the whole phase space on
condition when $\lambda\neq 1$. On the
other hand in case when $\lambda=1$ we
find that this Poisson bracket vanishes
for the subspace of the phase space
where $R_{ij}=0$. However when
$R_{ij}=0$ we see that $\Phi_2$ is
preserved during the time evolution of
the system and hence it is not
necessary to impose additional
constraint $\Phi_2^{II}\approx 0$. Then
 $\Phi_1^{II}\approx 0$ is the
first class constraint. The gauge
fixing of given constraints implies
that all metric components and their
conjugate momenta are non-propagating
degrees of freedom and hence the theory
on the subspace $R_{ij}=0$ is
topological with no local degrees of
freedom.}

.

We see that the requirement of the
preservation of the secondary
constraints does not imply additional
constraints so that we obtained
following constraint structure.
 We have first
 class constraints $\mH_i\approx
0, p_\nu\approx 0, p_i\approx
0,p_\mA\approx 0$ together with  two
global first class constraints
$p_N\approx 0, \bT\approx 0$. Then we
have two  second class constraints
$\Phi_1^{II},\Phi_2^{II}$. The detailed
discussion of these constraints will be
given in the next section.
In this section we performed the
Hamiltonian analysis of
non-relativistic covariant HL gravity
in the formulation presented in
\cite{daSilva:2010bm} and we showed
that it leads to the same structure of
the constraints as in the original
proposal \cite{Horava:2010zj}. Note
that our analysis is valid for general
$\lambda$ with agreement with
\cite{daSilva:2010bm}. Further, as was
shown in \cite{Horava:2010zj} the
number of physical degrees of freedom
is the same as in the General
Relativity even if the constraint
structures of these two theories are
different. On the other hand the
non-relativistic covariant HL gravity
has an additional global Hamiltonian
constraint. However when we consider
 RFDiff invariant HL gravity as the
 starting point for $U(1)$ extension of
 HL Gravity we find theory with the same content
 of physical degrees of freedom as
 in non-relativistic covariant HL gravity
 with additional important difference
 which is an absence of  the global Hamiltonian
 constraint.

\section{Non-Relativistic Covariant RFDiff HL
Gravity}\label{fourth}
 RFDiff invariant Ho\v{r}ava-Lifshitz
 gravity was introduced in
\cite{Blas:2010hb} and further studied in
\cite{Kluson:2010na}. This is the version
of the Ho\v{r}ava-Lifshitz gravity that
is  not invariant under
 foliation preserving diffeomorphism
 but only under reduced set of diffeomorphism
\begin{equation}\label{RFDtr}
t'=t+\delta t \ , \quad  \delta
t=\mathrm{const} \ , \quad  x'^i=
x^i+\xi^ i(t,\bx)
\end{equation}
 As was argued in  \cite{Kluson:2010na}
  the simplest
form of RFDiff invariant
Ho\v{r}ava-Lifshitz gravity takes the form
\begin{equation}\label{RFDiffaction}
S=\frac{1}{\kappa^2} \int dt d^D\bx
\sqrt{g}(\tK_{ij}
\mG^{ijkl}\tK_{kl}-\mV(g)) \ ,
\end{equation}
where
\begin{equation}
\tK_{ij}=\frac{1}{2}(\partial_t g_{ij}
-\nabla_i N_j-\nabla_j N_i) \ .
\end{equation}
Note that this action differs from
 HL gravity action
(\ref{actionNOJI}) by absence of the
lapse $N$ and by replacement of the
extrinsic curvature $K_{ij}$ with
$\tK_{ij}$ given above.  This action is
invariant under  RFDiff symmetries
(\ref{RFDtr}) that is reduced with
respect to foliation preserving
diffeomorphism.

In order to find the $U(1)$ extension
of given theory we    introduce the
field $\nu$ and replace $N_i$ with
$\hN_i$ as
\begin{equation}
\hN_i=N_i-\nabla_i \nu \ .
\end{equation}
Then it is again easy to see that
the action  is invariant under
transformation
\begin{equation}\label{U1RTF}
N_i'(t,\bx)=N_i(t,\bx)+ \nabla_i
\alpha(t,\bx) \ , \quad \nu'(t,\bx)=
\nu(t,\bx)+\alpha(t,\bx) \ .
\end{equation}
Clearly this replacement is as trivial
as the one performed in the projectable
version of Ho\v{r}ava-Lifshitz gravity.
Then following the same procedure as in
 section (\ref{second}) we find the
 action in  the form
\begin{equation}\label{RFDaction}
S_{RFD}=\frac{1}{\kappa^2} \int dt
d^D\bx \sqrt{g}(\hK_{ij}
\mG^{ijkl}\hK_{kl}-\mV(g)+\mG(R)(\mA-a))
\ ,
\end{equation}
where
\begin{equation}
\hK_{ij}=\frac{1}{2}
(\partial_t g_{ij}-\nabla_i N_j-
\nabla_j N_i+\nabla_i\nabla_j \nu+
\nabla_j\nabla_i\nu) \ .
\end{equation}
Clearly this action is invariant under
(\ref{RFDtr}) and under (\ref{U1RTF}).
Further
 $\mA$ and $a$ transform as scalar
under (\ref{RFDtr})
\begin{equation}
\mA'(t',\bx')=\mA(t,\bx) \ , \quad
a'(t',\bx')= a(t,\bx) \
\end{equation}
Note that the action (\ref{RFDaction})
can be derived from non-relativistic
covariant HL action by setting $N=1$
and hence  one can expect that
 these theories describe the same
 local physics. However the my difference
 between these two formulations
emerges when we perform the Hamiltonian
analysis of the
 action (\ref{RFDaction}).

  As in previous
section we find the
primary constraints
\begin{equation}
p_i(\bx)\approx 0 \ ,  \quad \Phi_1:
p_\mA(\bx)\approx 0 \ , \quad \Phi_2:
p_\nu(\bx)+\frac{1}{\kappa^2}\sqrt{g}
\mG(\bx) \
\end{equation}
and the relation between $\hK_{ij}$ and
conjugate momenta $\pi^{ij}$
\begin{equation}
\hK_{ij}=\frac{1}{\sqrt{g}}
\mG_{ijkl}\pi^{kl}\ .
\end{equation}
Then it is easy to find
the total Hamiltonian in the form
\begin{eqnarray}\label{HRFD}
H&=&\int d^D\bx (\mH_T+N^i\mH_i+
v^i\Phi_i+v_Np_N+v_i p^ i)-\nonumber \\
&-&\frac{1}{\kappa^2}
\int d^D\bx \sqrt{g}\mG(R)
(\mA-N^i \nabla_i \nu+\frac{1}{2}
\nabla^i \nabla_i\nu) \ ,
\nonumber \\
\end{eqnarray}
where
\begin{eqnarray}\label{mHTAr}
\mH_T&=&\frac{\kappa^2}{\sqrt{g}}\pi^{ij}\mG_{ijkl}\pi^{kl}
-\frac{1}{\kappa^2}
\sqrt{g}\mV(g)-\frac{2}{\kappa^2}\nu \nabla_i\nabla_j\pi^{ji} \ ,
\nonumber \\
\mH_i&=&-2g_{il} \nabla_k \pi^{kl} \ . \nonumber \\
\end{eqnarray}
The
requirement of the preservation of the
primary constraints $
 p_i(\bx)\approx 0 ,
\Phi_{1}(\bx)\approx 0$  implies following
secondary ones
\begin{eqnarray}
\partial_t \Phi_1&=&\pb{\Phi_1,H}=
-\frac{1}{\kappa^2}
\sqrt{g}\mG(R^{(D)})\equiv -
\Phi_1^{II}
\approx 0 \ ,
\nonumber \\
\partial_t p_i&=&\pb{p_i,H}=
-\mH_i-\Phi_1^{II} \approx
-\mH_i\approx 0
 \  .  \nonumber \\
\end{eqnarray}
In case of the preservation
of the constraint $\Phi_2$
we proceed as in previous section
and we find
 \begin{eqnarray}\label{tPhi2RF}
\partial_t \Phi_2&=&
\pb{\Phi_2,H}\approx - 2\frac{d\mG}{dR}
\left(
 R^{ij}
\mG_{ijkl}\pi^{kl} -
\frac{1-\lambda}{(\lambda
D-1)}\nabla_k
\nabla^k \pi\right)=\frac{d\mG}{dR}
\Phi_2^{II} \ ,
\nonumber \\
\end{eqnarray}
where
\begin{eqnarray}
\Phi_2^{II}
=- R_{ij}\pi^{ji}
 +\frac{\lambda}{D\lambda-1}R\pi
+ \frac{1-\lambda}{(\lambda
D-1)}\nabla_k \nabla^k \pi
\equiv
M_{ij}(g(\bx))\pi^{ji}(\bx)
\ ,  \nonumber \\
\end{eqnarray}
and where generally $M_{ij}(g(\bx))$ is   a
differential operator acting on $\pi^{ij}$
 that it reduces to ordinary multiplicative
operator in case $\lambda=1$. Note also
that $p_\nu\approx 0$ is the first class constraint.

Following general analysis of
constraints systems we introduce  the total
Hamiltonian in the form
\begin{eqnarray}
H_T&=&\int d^D\bx (\mH_T+N^i\mH_i+
v^{\mA}p_{\mA}+v^\nu
p_\nu+v_ip^i+v_{II}^1\Phi_1^{II}
+v_{II}^2\Phi_2^{II})\ . \nonumber \\
\end{eqnarray}
As the final step we should perform
the analysis of the secondary constraints.
However this was done in previous section
so that we do not repeat it here.

Let us now discuss the second class constraints
$\Phi_1^{II},\Phi_2^{II}$. According to standard
analysis these constraints have to vanish strongly
and allow to solve for two phase space variables
as a functions of remaining physical phase space
variables that span the reduced phase space.
However solving these
constraints in
full generality is very difficult. On the other hand
it is easy to see that in
linearized approximation these constraints
can be solved as $h=0\ , \pi=0$ where
$h$ is the trace part of the metric
fluctuation and $\pi$ is its conjugate momenta.

Even if we cannot solve these constraints
explicitly in general case we can still
determine the number of physical degrees
of freedom. To do this note that there are
$D(D+1)$  gravity phase   space variables   $g_{ij}, \pi^{ij}$, $2D$
variables $N_i,p^i$,
$2$ variables $\mA,p_\mA$
and $2$  variables $\nu,p_\nu$.
In summary the total number of degrees of freedom is
 $N_{D.o.f}=D^2+3D+4$.
 On the other
hand we have $D$  first class
constraints $\mH_i\approx 0$, $D$   first class
constraints  $p_i\approx 0 $, $2$   first class
constraints $p_\nu\approx ,p_\mA\approx 0$ and two
second class constraints $\Phi_1^{II},
\Phi_2^{II}$. Then we have $N_{f.c.c}=2D+2$
  first class constraints and
 $N_{s.c.c.}=2$  second
class constraints. Then the
number of physical degrees of freedom
is \cite{Henneaux:1992ig}
\begin{equation}\label{pdf}
N_{D.o.f.}-2N_{f.c.c}-N_{s.c.c.}=
D^2-D-2 \
\end{equation}
that exactly corresponds to the number
of the phase space  physical degrees of freedom of $D+1$ dimensional
gravity.
For example for  $D=3$ the equation (\ref{pdf}) is equal
 to  $4$ which is  the number
 of phase space degrees of freedom of
 massless graviton.

In  summary the Hamiltonian of non-relativistic general covariant
RFDiff HL gravity gives the appropriate
number of  physical degrees of freedom of
gravitational theory
without introducing
global Hamiltonian constraint.
There is also  another interesting
aspect of given theory which is its
non-trivial symplectic structure. In
fact, let us denote the constraints
$\Phi_{1,2}^{II}$ as $\Phi_A^{II}$
where $A,B=I,II$ so that the Poisson
bracket between constraints can be
written as
\begin{equation}
\pb{\Phi^{II}_A(\bx),\Phi^{II}_B(\by)}=
\triangle_{AB}(\bx,\by) \ .
\end{equation}
From the structure of these constraints we find
that the matrix $\triangle_{AB}$ has
following structure
\begin{equation}
\triangle_{AB}(\bx,\by)=
\left(\begin{array}{cc}
0 & * \\
 \ * & * \\ \end{array}\right) \ ,
\end{equation}
where $*$ means non-zero elements.
Then the inverse matrix
$(\triangle^{-1})_{AB}$ has the form
\begin{equation}\label{triin}
(\triangle^{-1})^{AB}=
\left(\begin{array}{cc}
* & * \\
\ * & 0 \\ \end{array}\right) \ .
\end{equation}
  Now we observe
that
\begin{eqnarray}\label{gcon}
\pb{g_{ij}(\bx),\Phi_1^{II}(\by)}=0 \ ,
\pb{g_{ij}(\bx),\Phi_2^{II}(\by)}\neq 0 \ ,
\nonumber \\
\pb{\pi^{ij}(\bx),\Phi_1^{II}(\by)}\neq 0 \ ,
\pb{\pi^{ij}(\bx),\Phi_2^{II}(\by)}\neq 0 \ .
\nonumber \\
\end{eqnarray}
Then we find that the Dirac brackets
between canonical variables take the form
\begin{eqnarray}
& &\pb{g_{ij}(\bx),g_{kl}(\by)}_D=
-\int d\bz d\bz'
\pb{g_{ij}(\bx),\Phi_A^{II}(\bz)}
(\triangle^{-1})^{AB}(\bz,\bz')
\pb{\Phi_B^{II}(\bz'),g_{kl}(\by)}=0 \ ,
\nonumber \\
& &\pb{\pi^{ij}(\bx),\pi^{kl}(\by)}_D=
\nonumber \\
&=&-\int d\bz d\bz'
\pb{\pi^{ij}(\bx),\Phi_A^{II}(\bz)}
(\triangle^{-1})^{AB}(\bz,\bz')
\pb{\Phi_B^{II}(\bz'),\pi^{kl}(\by)}=
\Omega^{ijkl}(\bx,\by) \ ,
\nonumber \\
& &\pb{g_{ij}(\bx),\pi^{kl}(\by)}_D=
\pb{g_{ij}(\bx),\pi^{kl}(\by)}
-\nonumber \\
&-&\int d\bz d\bz'
\pb{g_{ij}(\bx),\Phi_A^{II}(\bz)}
(\triangle^{-1})^{AB}(\bz,\bz')
\pb{\Phi_B^{II}(\bz'),\pi^{kl}(\by)}=
\Omega_{ij}^{kl}(\bx,\by) \ ,
\nonumber \\
\end{eqnarray}
where the matrix $\Omega$ depends on
phase-space variables according to
(\ref{triin}) and (\ref{gcon}). Hence
the  non-relativistic covariant RFDiff
 HL gravity has well defined
Hamiltonian formulation with
symplectic structure that generally
depends on phase space variables. Note however
that in case of the linearized approximation
one can choose the constraints in such
a way that the Dirac bracket coincides
with the Poisson bracket. Explicitly,
in linearized approximation
the second class constraints can be
chosen as $h=0,\pi=0$ as follows from the
analysis given above. These constraints
 have vanishing Poisson brackets
with remaining dynamical variables and
consequently the Dirac brackets between
physical phase space variables  coincide
with Poisson brackets.

As the final remark we again emphasize
the important point that  the Hamiltonian
of non-relativistic covariant RFDiff HL gravity
 does not
vanish on constraint surface.
This is the similar situation  as in case
of the Hamiltonian of the healthy
extended HL gravity
\cite{Kluson:2010xx,Kluson:2010nf}
which  is however in sharp contrast  with
the Hamiltonian of General Relativity. As
we argued in these papers this fact has
a strong impact on the definition of
observables in healthy extended Ho\v{r}ava-
Lifshitz gravity or in any theory of gravity
where the Hamiltonian is not given as
linear combination of constraints.
Since the discussion presented in
\cite{Kluson:2010xx,Kluson:2010nf}
can be applied in case of the non-relativistic
covariant RFDiff HL gravity as well we are not going
to repeat it here. Instead we recommend
these papers to reader that is interested in
these problems.

 It is important to stress that the fact that
 the
Hamiltonian is not given as linear
combination of constraints
 has an
 important consequence for the
 stability of given theory. Explicitly,
 it is well known that some massive
 gravities are unstable since the
 Hamiltonian is not bounded from
 bellow. Alternatively, the instability
 of given theory is also indicated by
 presence of the ghosts (fields with
 wrong sign of kinetic term) in the
fluctuation spectrum. However we
believe that this is not the case of
non-relativistic covariant RFDiff
 HL gravity even if the full
analysis has not been done yet. The
crucial fact is the absence of the
scalar graviton in the fluctuation
spectrum that implies that RFDiff
invariant HL gravity has well defined
Hamiltonian that is positive definite
at least in linearized approximation.
In order to understand the properties
of the Hamiltonian of non-linear theory
we should solve the second class
constraints and express the Hamiltonian
in terms of physical modes only.
However as we argued above this is very
difficult task and hence the problem of
the stability of the general
non-relativistic covariant RFDiff HL
gravity has not been fully addressed.
 \noindent {\bf
Acknowledgements:}

I would like to
thank to A. Kobakhidze
 for very useful discussions
 that led to important corrections in
 the first version of my paper.
 This work was also supported by the
Czech Ministry of Education under
Contract No. MSM 0021622409. I would
like also thank to Max Planck Institute
at Golm for its financial support and
 kind hospitality during my work on
this project.

\end{document}